\documentclass[ALICE,manyauthors]{cernphprep}
\usepackage[comma,square,numbers,sort&compress]{natbib}
\usepackage[T1]{fontenc} 
\usepackage{hyperref}
\usepackage{lineno}
\usepackage{multicol}
\usepackage{multirow}
\usepackage{xcolor}

\usepackage{siunitx}
\usepackage{orcidlink}

\newcommand{\dndeta}       {\ensuremath{\mathrm{d}N_\mathrm{ch}/\mathrm{d}\eta}}
\newcommand{\pT}           {\ensuremath{p_\mathrm{T}}}
\newcommand{\avdndeta}{\langle \mathrm{d}N_\mathrm{ch} / \mathrm{d}\eta \rangle}
\newcommand{\cmss}{$\sqrt{s}=13.6$~\text{TeV}}
\newcommand{\inelg}{INEL\textgreater0\xspace}

\begin{document}%

\begin{titlepage}
\PHyear{2026}       
\PHnumber{012}      
\PHdate{21 Jan}  

 \title{Inclusive and multiplicity-dependent pseudorapidity densities of charged particles in pp collisions at $\mathbf{\sqrt{s} = 13.6}$~TeV}
\ShortTitle{Pseudorapidity densities in pp collisions at $\mathrm{\sqrt{s} = 13.6}$~TeV}  

\Collaboration{ALICE Collaboration\thanks{See Appendix~\ref{app:collab} for the list of collaboration members}}
\ShortAuthor{ALICE Collaboration} 

\begin{abstract}
The distribution of the pseudorapidity density of primary charged particles ($\dndeta$) in $|\eta|<1.5$ is measured in pp collisions at a center-of-mass energy of 13.6 TeV using the ALICE detector. Tracks are reconstructed with the upgraded Inner Tracking System and Time Projection Chamber, and results are reported for events having at least one charged particle in $|\eta|<1$. The multiplicity dependence of the pseudorapidity density of charged particles is also explored with the multiplicity determined by the forward detectors.

The average charged-particle pseudorapidity density at midrapidity ($|\eta|<0.5$) is measured to be $\langle \mathrm{d}N_{\mathrm{ch}}/\mathrm{d}\eta \rangle = 7.10 \pm 0.20$. Combined with lower-energy measurements, the results follow a power-law scaling with the center-of-mass energy. 
These measurements provide a new reference for charged-particle production at the highest proton–proton collision energy available at the LHC.

\end{abstract}
\end{titlepage}

\setcounter{page}{2} 


\section{Introduction}

The multiplicity of primary charged particles is an important observable in high-energy hadronic collisions. This observable provides information about soft and hard partonic scattering processes~\cite{Field:1976ve}. The pseudorapidity ($\eta$) density of charged particles, $\mathrm{d}N_\mathrm{ch}/\mathrm{d}\eta$, was previously studied by ALICE and other LHC experiments across energies ranging from 0.9 to 13 TeV~\cite{ALICE:2009wpl,ALICE:2010cin,ALICE:2010mty,ALICE:2015olq,ALICE:2015qqj,ALICE:2020swj,ALICE:2022kol,CMS:2015zrm,ATLAS:2016zkp}, and it offers insights into the interplay between perturbative and non-perturbative QCD particle production mechanisms. Typically, particle production is dominated by soft processes, but it also includes substantial contributions from hard scattering. While these measurements constrain the global features of proton--proton (pp) collisions, they do not reveal how the production mechanism evolves with event activity, especially in high-multiplicity events.

Studies of high-multiplicity events in pp and proton--nucleus (pA) collisions at the LHC have shown similar effects to those observed in nucleus--nucleus (AA) collisions in the soft regime. 
One such phenomenon is the ridge structure in long-range azimuthal correlations in pp and p--Pb collisions~\cite{CMS:2010ifv,ALICE:2012eyl, ALICE:2022wpn}.
Additionally, ALICE reported enhanced production of (multi-) strange particles with increasing multiplicity in pp collisions~\cite{ALICE:2016fzo, ALICE:2022wpn}, suggesting a strong correlation with event activity, which is characterized by the final-state charged-particle multiplicity.
These similarities can be interpreted as potential signatures of quark--gluon plasma (QGP) formation. 

The event multiplicity in pp and pA collisions is closely related to the number of Multiple-Parton Interactions (MPIs), a key component of particle production models. Higher multiplicity events typically result from an increased number of MPIs, leading to enhanced particle production~\cite{ALICE:2013tla,ALICE:2014mas}. 
The color reconnection (CR) mechanism~\cite{Christiansen:2015yqa, OrtizVelasquez:2013ofg} is an essential feature for MPI modeling in PYTHIA~8~\cite{Skands:2014pea, Sjostrand:2014zea, Bierlich:2022pfr} and it is able to reproduce some of the features observed in small systems for high multiplicity events. On the other hand, EPOS~\cite{Werner:2023zvo} controls MPI through energy sharing and saturation within a Gribov–Regge multiple-scattering scheme, and by implementing a core–corona mechanism in which high string-density ``core'' regions hadronize collectively while low-density ``corona'' regions fragment as strings. 
This approach allows EPOS to reproduce flow-like behavior and strangeness enhancement observed in small collision systems.
Therefore, measurements of multiplicity-dependent pseudorapidity densities provide valuable constraints on MPI modeling~\cite{ALICE:2020swj}.

In the event generators considered here, MPI modeling affects high-multiplicity events by shaping correlations between transverse momentum ($p_\mathrm{T}$) of particles and multiplicity, pseudorapidity distributions, and  $p_\mathrm{T}$ spectra~\cite{ATLAS:2010jvh, Gieseke:2012ft, Kaidalov:1982xg}. This work presents the multiplicity dependence of the pseudorapidity density in pp collisions. Motivated by these considerations, this work presents inclusive and multiplicity-differential measurements of $\dndeta$ of primary charged particles in pp collisions at $\sqrt{s}=13.6$~TeV, thus extending previous ALICE $\dndeta$ measurements to the highest pp energy to date and providing constraints for MPI and particle-production modeling. 

This article is organized as follows: Sec.2 describes the experimental setup and data collection procedures; Sec.3 explains the analysis procedure and corrections; Sec.4 presents a detailed evaluation of systematic uncertainties of the analysis. Sec.5 reports the results and discusses the implications of these findings for MPI modeling, event generator tuning, and QCD phenomenology. Lastly, Sec.6 provides a summary and conclusions. 

\section{Experimental setup}

The ALICE collaboration carried out a major upgrade of the ALICE detector during the LHC Long Shutdown 2 (2019--2022) to prepare for Run 3 operations~\cite{ALICE:2022wpn}.
The key changes include the transition to continuous readout and the introduction of a new online event processing integrated with a redesigned online-offline software framework~\cite{Buncic:2011297}. 
These significantly improved readout capabilities allow for precise and differential characterization of the QGP phase and for exploration of previously inaccessible rare signals.

The experimental setup consists of a central barrel housed inside a solenoidal magnet providing a magnetic field of up to 0.5 T. The central barrel is designed for efficient track reconstruction in a high-density track environment of heavy-ion collisions, covering transverse momenta from 100 MeV/$c$ to 100 GeV/$c$ with excellent capabilities for identifying hadrons and electrons. For the track selection used in this analysis, the tracking efficiency is approximately 82\% at $p_{\mathrm T}=100$ MeV/$c$ and remains close to 90\% from approximately 1 to 100 GeV/$c$. 

One of the main upgrades in Run 3 is the new ITS, which employs seven layers of ALPIDE monolithic active pixel sensors (MAPS)~\cite{ALICE:2013nwm}. The sensors offer enhanced pointing resolution due to their reduced material budget, proximity to the interaction point and improved spatial resolution compared to the original ITS used during LHC Runs 1 and 2. The new ITS ensures a robust performance in high interaction rates and high-multiplicity conditions, where multiple collisions may occur within the same ITS readout frame. 
The $p_{\mathrm T}$-integrated tracking efficiency is approximately 87\%, with a variation of less than 1\% across the multiplicity classes considered in this analysis.

The TPC is located radially outside the ITS, spanning a radius from 0.85 m to 2.5 m and a total length of 5 m. For Run 3, the TPC substituted the previously used multi-wire proportional chambers with Gas Electron Multiplier (GEM) foils~\cite{CERN-LHCC-2013-020}. 
The GEMs allow for continuous readout while keeping the ion backflow under control, reducing space-charge effects to a level that can be corrected online, with further refinements applied offline to achieve physics-ready calibration.

The Fast Interaction Trigger (FIT) system is used for event selection and timing~\cite{Antonioli:1603472}. FIT comprises two arrays of fast Cherenkov radiators (FT0) positioned on either side of the interaction point, along with three sets of scintillator detectors. 
The FT0 covers $3.5<\eta<4.9$ (FT0A) and $-3.3<\eta<-2.1$ (FT0C), and it provides the collision time.

\section{Data sample and analysis}

The beginning of the LHC Run 3 in 2022 was marked by achieving a center-of-mass energy of $\sqrt{s} = $13.6 TeV in pp collisions. Since then, the ALICE experiment collected a dataset more than 300 times larger than that recorded during the Runs 1 and 2 with a minimum-bias trigger. In this article, in order to study the charged-particle pseudorapidity density, a subset (3.4 billion events, $\mathcal{L}_{\mathrm{int}}=57.05~\mathrm{nb}^{-1}$) of the 58 billion events collected in 2022 is analyzed. 

The results of this paper are presented for the \inelg event class, which is defined as inelastic events with at least one charged particle at midrapidity ($\left|\eta\right| < 1$). 
This event class largely suppresses diffractive contributions and therefore does not require explicit corrections for diffractive events. It has been widely adopted as a common convention across LHC experiments during Runs~1 and~2.
At sufficiently large collision energy, particles produced in diffractive events almost always have enough boost to completely miss central detectors~\cite{ALICE:2015olq}. At the time of this publication, no measurement of the diffractive cross section at $\sqrt{s} = 13.6$ TeV is available. Since the measurement of $\dndeta$ in inelastic or non-single diffractive collisions would require reliable modeling of the diffractive component--both its cross section and diffractive mass spectrum, such normalization would introduce significant model dependence. The normalization to the INEL>0 class allows us to minimize the diffractive contribution, as most of the diffractive events would not yield particle production within $|\eta|<1$. The residual contamination is estimated to be negligible~\cite{ALICE:2015olq,ALICE:2012fjm}.

The selection of minimum-bias (MB) events is based on several criteria to ensure high-quality collision samples. The primary requirement is a vertex trigger coincidence, where the reconstructed vertex must be temporally aligned with a bunch crossing, as determined by the corresponding FT0 detector signal. To prevent the inclusion of partially recorded events, collisions that occur near the ITS readout frame borders or the time frame borders are excluded. The ITS readout frame borders correspond to the edges of the detector's local readout window, where hits may be truncated due to a limited integration time. 
Similarly, the time-frame borders correspond to the boundaries of the global continuous readout intervals (10--20 ms). Near the start and end of each time frame, a small fraction of data --- about one TPC drift time ($\sim$\SI{90}{\micro\second}) --- may suffer from incomplete charge collection, as signals from these regions are not fully contained within the same frame. To avoid partially reconstructed events, collisions within these intervals are excluded from the analysis.
Additionally, events are required to have a reconstructed longitudinal vertex position within $\pm10$~cm of the nominal interaction point. 
This selection ensures a uniform detector acceptance and reconstruction efficiency across the pseudorapidity range, consistent with the approximately Gaussian vertex distribution with a sigma of about 5~cm. This selection removes approximately 3\% of the events.

Further quality criteria are applied to ensure the reliability of the reconstructed event sample. Events in which the primary vertex has no associated global tracks, defined as tracks reconstructed using both ITS and TPC hits, are removed. Additionally, events where the FT0-estimated vertex position deviates by more than 1 cm from the reconstructed vertex position are excluded.

For multiplicity classification, the FT0M forward multiplicity estimator is used, defined as the sum of FT0 signal amplitudes from both the A and C sides of the detector. The multiplicity percentiles are calibrated to ensure a consistent total visible cross section fraction across different data-taking periods, providing a uniform event classification. Multiplicity classes are defined based on percentile ranges of the total visible cross section in minimum-bias pp collisions with at least one track in $|\eta|<1$ (MB>0), where 100\% corresponds to the entire visible cross section. Higher multiplicity events are associated with lower percentile values (closer to 0\%), while lower multiplicity events correspond to higher percentile values (closer to 100\%). It is worth noting that the event activity classes quoted in this paper (e.g. MB>0 percentile 0--1\%, 1--5\%, …) are defined as detector-level percentiles of the FT0M estimator in the MB>0 event samples, and do not correspond to percentiles of the INEL>0 cross section.
This multiplicity definition is similar to the Run~2 V0M classification, which relied on the combined amplitudes of the V0A ($2.8<\eta<5.1$) and V0C ($-3.7<\eta<-1.7$) detectors~\cite{ALICE:2013axi}.
The FT0M estimator is associated with the upgraded FIT system ($3.5<\eta<4.9$ and $-3.3<\eta<-2.1$), which features superior timing and rate performance necessary for high-luminosity operation.

The multiplicity percentile intervals defined with the 
visible cross section measured using the MB>0 selection can be 
translated into fractional intervals with respect to the INEL>0 cross section. 
This conversion is performed by applying

\begin{equation}
P_i(\text{INEL>0}) =
\frac{P_i(\text{MB>0}) / \varepsilon_i}
     {\sum_{j} \left[ P_j(\text{MB>0}) / \varepsilon_j \right]},
\end{equation}

where $i$ denotes a given multiplicity class and the index $j$ runs over all 
multiplicity intervals.  
The quantity $\varepsilon_i$ represents the MB>0 trigger 
efficiency for selecting INEL>0 events in class $i$. The multiplicity percentile interval
$P_i(\text{MB>0})$ corresponds to the fraction of events fulfilling the 
detector-level MB>0 condition, while the normalized expression above provides the corresponding fraction relative to the 
INEL>0 event sample. For Run~3, ALICE multiplicity classes are defined 
in terms of $P(\text{MB>0})$, which is a detector-level observable. 

The transition to the continuous readout data acquisition system requires modifications in the event reconstruction strategy. With the triggered readout approach, the association of reconstructed tracks to a collision vertex was straightforward, with the only ambiguity arising from the same-bunch pileup. However, in continuous readout mode, track-to-vertex association relies on track timestamps matched to the FT0 detector vertex signals, making it dependent on the system’s time resolution.

The precision of the track timestamp depends on the intrinsic time resolution of the contributing detectors and the number of timing measurements. Tracks with large time uncertainties, or those originating from secondary decays (with $c\tau$ of the order of millimeters to centimeters), may be compatible with more than one reconstructed vertex, particularly in high pile-up scenarios. However, dedicated studies indicate that the timestamp-based association to the most probable vertex remains robust. The fraction of tracks that are more compatible with an alternative vertex does not exceed the percent level up to interaction rates of about 1 MHz from high to low multiplicity events~\cite{Antonioli:1603472, Buncic:2011297}. Thus, the impact of track-to-vertex association ambiguity is considered negligible for the present analysis.

\begin{figure}[h]
\centering
\includegraphics[width=0.8\linewidth]{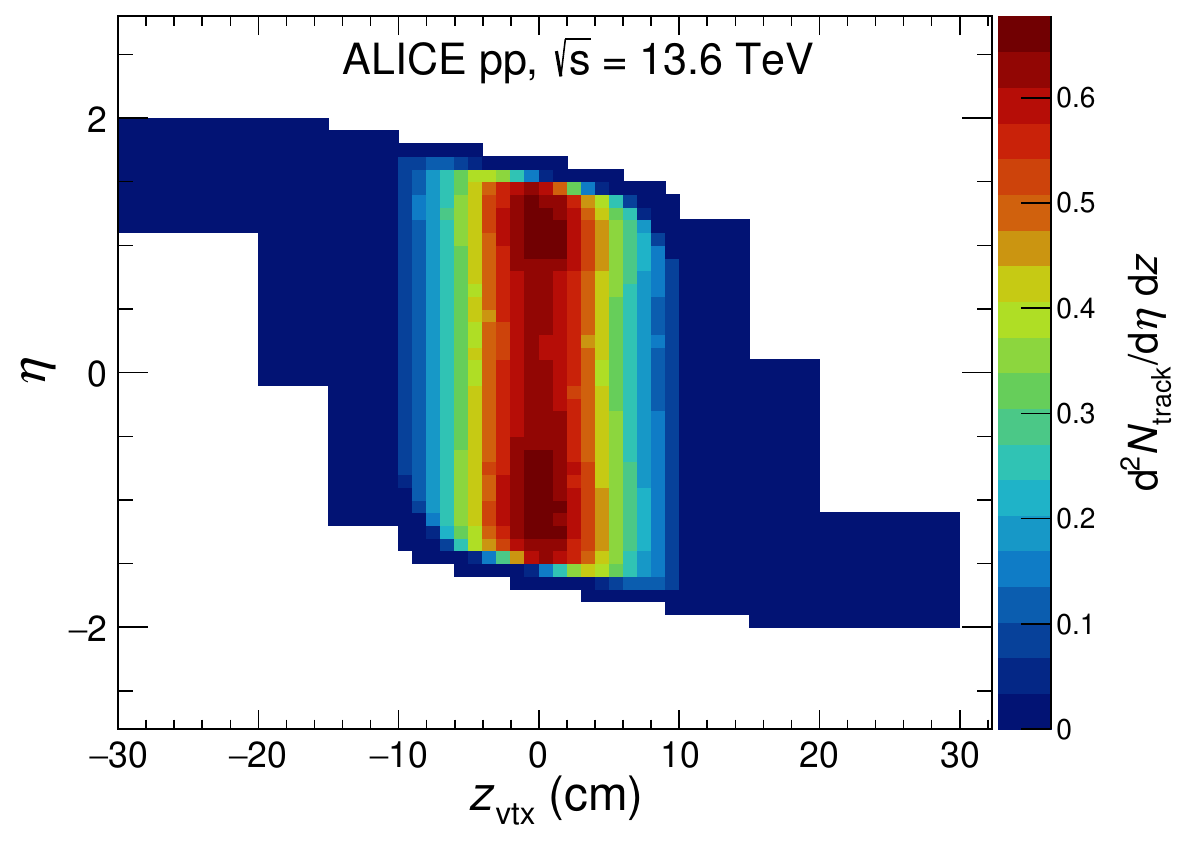}
\caption{Event-normalized distribution of reconstructed tracks in the $(z_{\mathrm{vtx}},\eta)$ plane for pp collisions at $\sqrt{s}=13.6$~TeV. The quantity $\mathrm{d}^{2}N_{\mathrm{track}}/\mathrm{d}\eta\,\mathrm{d}z$ is shown as a function of the reconstructed primary-vertex position $z_{\mathrm{vtx}}$ and pseudorapidity $\eta$. 
}
	\label{fig:d2ndzdeta}
\end{figure}

Tracking in the pseudorapidity range covered in this study is performed using two sub-detectors: ITS and TPC. For events with a vertex at the nominal interaction point ($z_\mathrm{vtx}=0$), the geometrical acceptance of the TPC extends up to $|\eta|<0.9$, while the ITS provides tracking coverage up to $|\eta|<1.22$. In this analysis, due to the large available data sample, the inclusion of events with displaced vertices allows the extension of the pseudorapidity coverage to $|\eta|<1.5$. 
The dependence of the pseudorapidity acceptance on the reconstructed vertex position is illustrated in Fig.~\ref{fig:d2ndzdeta}, which shows the event-normalized distribution of reconstructed tracks in the $(z_{\mathrm{vtx}},\eta)$ plane. The figure makes it explicit that the outermost $\eta$ intervals are populated mainly by events with large $|z_{\mathrm{vtx}}|$, while the central region receives contributions over the full accepted vertex range. 

Reconstructed tracks can be categorized into three exclusive classes based on the contributing detectors: global tracks, which utilize both ITS and TPC measurements; ITS-only tracks, reconstructed solely from ITS hits; and TPC-only tracks, built exclusively from TPC measurements. Due to their insufficient pointing resolution at the vertex, a large uncertainty in the track-to-collision association, and the significant contribution of secondary particles, TPC-only tracks are not included in this analysis. The selected track sample consists of global tracks that pass quality criteria for both ITS and TPC contributions, as well as ITS-only tracks that meet the required quality conditions for ITS measurements. Both track categories must satisfy the requirements on a distance of closest approach (DCA) to the primary vertex ($\mathrm{DCA}_\mathrm{z}<2$~cm in the longitudinal direction and smaller than
$\mathrm{DCA}_{\mathrm{xy}} < 0.0105 + 0.035\,(p_{\mathrm{T}}/(\mathrm{GeV}/c))^{-1.1}$~cm
in the transverse plane) to ensure the selection of tracks originating from primary particles. Primary particles are defined as prompt particles produced in the collision with a proper lifetime of 
$c\tau > 1$~cm, as well as those originating from the decays of particles with shorter proper lifetimes, excluding those from weak decays of strange hadrons.

Three different MC corrections are applied to the raw $\dndeta$: 
(a) a track-to-particle correction that accounts for detector inefficiencies and background particles like secondaries from interactions in the detector material and decays of primary charged particles, including the strange particle content correction, (b) a vertex reconstruction efficiency correction for selected events without a reconstructed vertex, and (c) a selection efficiency correction, which accounts for the bias due to the selection requirement for the corresponding event class. All the track-to-particle corrections are applied as a function of $z_\mathrm{vtx}$ in order to consider the $z$-dependent acceptance and reconstruction efficiency of charged-particle tracks.

The ALICE definition of primary charged particles excludes particles originating from weak decays of strange particles. Therefore, data must be corrected for cases when daughter particles from these decays pass the track selection. The strangeness content in data was 40--50\% larger than predicted by PYTHIA 8~\cite{Sjostrand:2007gs} in previous analyses~\cite{ALICE:2016fzo}. 
However, the recent version of PYTHIA 8 (v.8.311) describes the content better, and the discrepancy remains within 20\%. The discrepancy is accounted for by scaling the strangeness content in the MC simulation to that in the data. The scaling factor is determined by measuring the ratio between the abundance of decay products in $|\eta|<0.8$ from reconstructed $\mathrm{K}^0_\mathrm{s}$, $\Lambda$, and $\bar{\Lambda}$ in data and MC simulations~\cite{ALICE:2015olq}. The scaling factor in the correction procedure compensates for the underestimated strangeness content in MC and results in a relative downward correction of less than 0.5\% on the final \dndeta~\cite{ALICE:2022kol}.

\section{Systematic uncertainties}

\begin{table}[h!tbp]
\centering
\caption{Relative systematic uncertainties (\%) for \dndeta\ in $|\eta|<0.5$. Note that specific multiplicity classes are merged to focus on the systematic uncertainty trend.}
\label{tab:systematic}
\small
  \resizebox{0.95\textwidth}{!}{%
\begin{tabular*}{\textwidth}{@{\extracolsep{\fill}} c | c  c  c  c  c  c  c  c  }
\hline

\multirow{2}{*}{Source of uncertainty}  & \multicolumn{8}{c}{Systematic uncertainty for multiplicity classes in $|\eta|<0.5$ (\%)} \\\cline{2-9}
                          &0--1\%&1--5\%&5--10\%&10--30\%&30--50\%&50--70\%&70--100\%&0--100\%\\
\hline 
Azimuthal acceptance        & 0.7 & 0.5 & 0.5 & 0.5 & 0.5 & 0.5 & 0.5 & 0.5 \\
$z_\mathrm{vtx}$ acceptance & 0.3 & 0.2 & 0.2 & 0.2 & 0.2 & 0.2 & 0.3 & 0.1\\
\pT\ extrapolation          & 1.1 & 1.1 & 1.1 & 1.2 & 1.2 & 1.3 & 1.3 & 2.4 \\
Strangeness content         &$<0.1$& 0.1 & 0.1 & 0.1 & 0.2 & 0.2 & 0.4 & 0.2 \\
Particle composition        & 0.5 & 0.5 & 0.5 & 0.5 & 0.5 & 0.5 & 0.5 & 0.5 \\

Material Budget             & 0.7 & 0.3 & 0.2 & 0.1 & 0.1 & 0.1 & 0.1 & 0.1 \\
Track selection             &$<0.1$&$<0.1$&$<0.1$&$<0.1$&$<0.1$&$<0.1$&$<0.1$&$<0.1$\\
Diffraction                 &$<0.1$&$<0.1$&$<0.1$&$<0.1$&$<0.1$&$<0.1$&0.1&0.2\\
Event and signal losses     & 0.2 & 0.2 & 0.2 & 0.1 & 0.2 & 0.4 & 0.8 & 1.1 \\
\hline
Total                       & 1.8 & 1.4 & 1.4 & 1.4 & 1.5 & 1.6 & 1.8 & 2.8 \\
\hline

\end{tabular*}
}
\end{table}

The contributions from various sources of systematic uncertainties associated with the measurement of the charged-particle pseudorapidity density are summarized in Tab.~\ref{tab:systematic}. 

The uncertainty due to acceptance and efficiency arises from the azimuthal dependence of the acceptance and the reconstruction efficiency of the detector. 
To estimate this effect, the charged-particle pseudorapidity density is measured separately in three different azimuthal sectors, and the results are compared to the measurement obtained using the full azimuthal coverage, which amounts to about 0.5--0.7\% depending on the multiplicity class. 
Another source of uncertainty comes from the dependence on the interaction vertex position along the beam direction ($z_\mathrm{vtx}$). To evaluate this effect, the analysis is performed in different slices of the $z_\mathrm{vtx}$ ranges and the results are compared to those obtained using the nominal vertex selection of $|z_\mathrm{vtx}|<10$~cm. The uncertainty associated with interaction-vertex position ranges from 0.1\% to 0.3\%. 

At low transverse momenta, the reconstruction efficiency for charged particles decreases significantly due to their deflection in the magnetic field and interactions such as scattering and absorption within the detector material. This effect is particularly pronounced below 100 MeV/$c$, where undetected low-\pT\ particles can impact the final measurement. To assess this uncertainty, the production rate of primary charged particles with $\pT<100$ MeV/$c$ is varied within the event generator by a constant factor of $+100$\% and $-50$\%. 
The resulting variation in the final $\dndeta$ amounts to 1.1–1.3\% depending on the multiplicity class, and 2.4\% for the inclusive one. 

The uncertainty associated with the strangeness reweighting correction is evaluated by varying the scaling factor applied to match the strangeness content in data and simulation. The factor, obtained from data-to-MC ratios of reconstructed $\mathrm{K}^0_\mathrm{s}$, $\Lambda$, and $\bar{\Lambda}$ yields, is varied by $\pm$30\% based on data-driven studies.The resulting variation in \dndeta\ is found to range from 0.4\% at low multiplicities to negligible at the highest multiplicity. 

The determination of the reconstruction efficiency is influenced by the composition of the produced particle species, as different hadrons exhibit varying detection efficiencies due to differences in their effective \pT\ thresholds and decay kinematics. 
To estimate the uncertainty associated with particle-species composition, the relative abundances of charged kaons, protons, and all remaining charged hadrons are varied by $\pm 30 \%$ with respect to the fixed number of charged pions in the simulation. 
This variation accounts for uncertainties in the measured particle composition at the LHC. The resulting impact on the final measurement is found to be 0.5\% across all multiplicity classes. 

Uncertainty in the detector material affects the determined \dndeta\ through secondary particles, mainly photon conversions ($\gamma\rightarrow\mathrm{e^+e^-}$) and hadronic interactions that create or absorb secondary particles and thereby influence their reconstruction efficiency.
Material-related effects to \dndeta\ are evaluated by repeating the correction with simulations in which the detector material is varied by $\pm10\%$ and compared with the nominal analysis. For each $\eta$ bin, the ratios of the $\pm10\%$-varied spectra to the nominal one are measured, and the material-budget uncertainty is obtained as the average deviation over the analyzed pseudorapidity region. The uncertainty decreases from high to low multiplicity, from $\sim$0.7\% in the 0--1\% class to $\sim$0.1\% in the 70--100\% class. 

The systematic uncertainty related to the track selection is evaluated by modifying the track-matching criteria in the ITS detector. In the baseline analysis, a track is required to have at least one hit in the first three ITS layers. To assess the impact of this selection, four alternative scenarios are considered:
(a) a stricter condition where a hit is required in each of the seven ITS layers, reducing tracking efficiency but improving track quality; (b) removing the DCA criteria, which increases the contribution of secondaries and mis-reconstructed tracks, to evaluate its influence on the measured $\dndeta$; (c) relaxing the TPC track quality requirements, such as the number of clusters ($>70$) and the $\chi^2$ per cluster ($<4$), to assess the effect of including lower-quality tracks; and (d) removing the ITS quality criteria for ITS and ITS-TPC (tracks reconstructed using the combined information of ITS and TPC) tracks, thereby ignoring minimum requirements on the cluster count and fit quality.

The contribution from diffractive processes is largely suppressed by the INEL>0 event selection. A generator-level study with PYTHIA 8 at $\sqrt{s}=13.6$ TeV predicts a residual diffractive fraction of about 4.8\% at midrapidity for the \inelg class. To estimate the associated model uncertainty, the fractions of single- and double-diffractive processes in the MC simulation are varied by $\pm30\%$ with respect to their nominal PYTHIA 8 values, and the full analysis correction is repeated for each variation. The RMS deviation from the nominal corrected \dndeta\ distribution is taken as the systematic uncertainty. The resulting uncertainty is negligible, remaining below 0.1\% for most multiplicity classes, reaching 0.1\% in the 70--100\% class, and amounting to 0.2\% for the inclusive 0--100\% result.

An additional uncertainty associated with event selection is evaluated by varying the simulated selection efficiency by a relative $\pm5$\% in each multiplicity class, motivated by indications of data–MC differences in the selection efficiency. The additional relative uncertainty ranges from 0.13\% to 0.80\% across the multiplicity classes and is 1.12\% for the inclusive measurement. 
Because the correction factors depend on multiplicity, correcting the integrated sample can respond differently to systematic variations than combining individually corrected classes. This can lead to larger inclusive uncertainties, as observed for the low-$p_{\mathrm T}$ extrapolation, diffraction, and event- and signal-loss contributions. The total systematic uncertainty is obtained by summing all individual contributions in quadrature.

\section{Results}

\begin{figure}[h]
\centering
\includegraphics[width=0.6\linewidth]{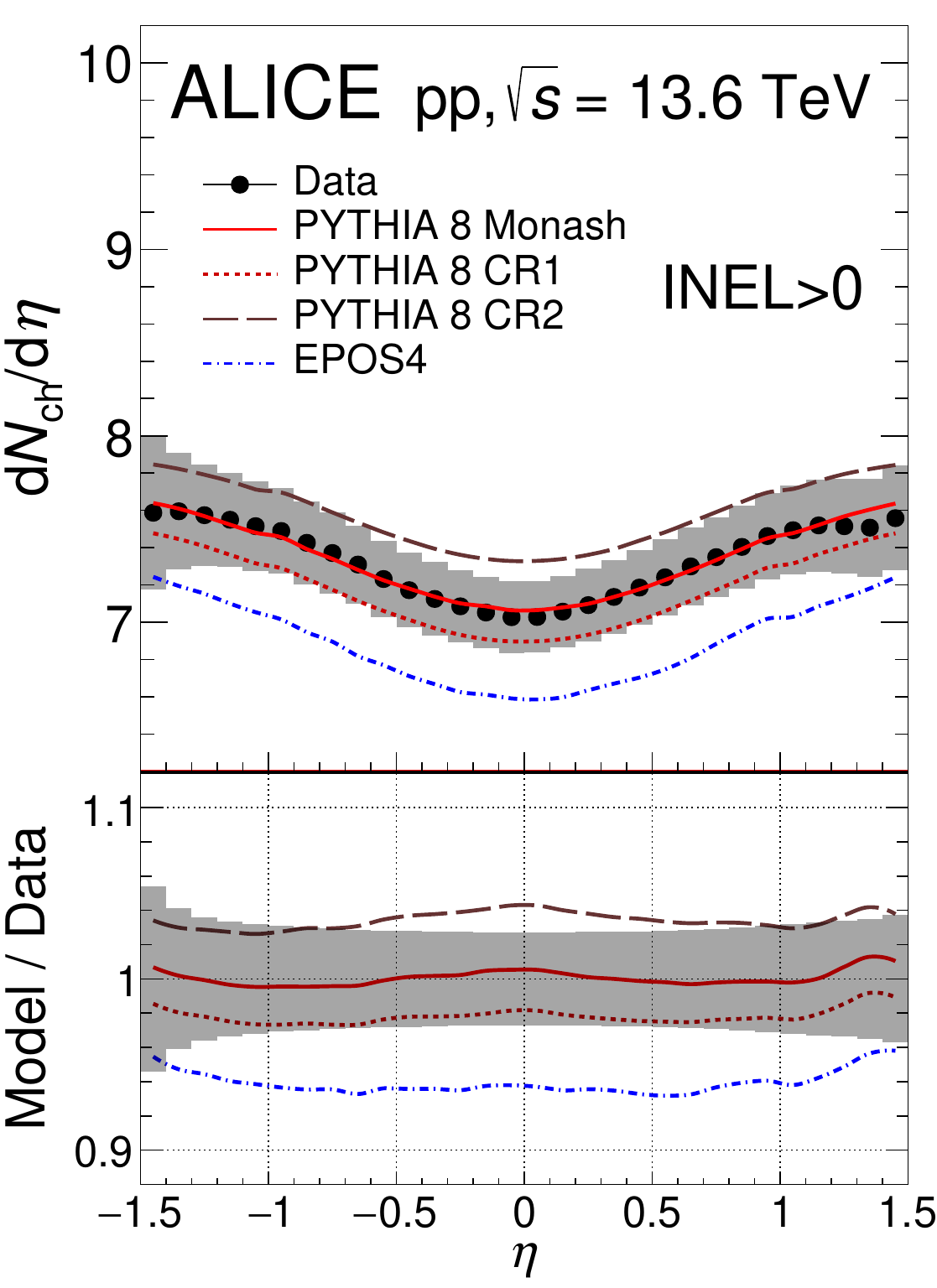}
\caption{The pseudorapidity distribution of primary charged particles for \inelg events in pp collisions at $\sqrt{s} = 13.6$ TeV. Data are compared to model predictions from PYTHIA 8 Monash 2013 and EPOS4. The systematic uncertainties in the data are represented by gray bands. The lower panel shows the ratio of model predictions to the data. }
	\label{fig:mbresult}
\end{figure}

The measurement of charged-particle pseudorapidity density (\dndeta) as a function of $\eta$ for \inelg\ events in pp collisions at $\sqrt{s} = 13.6$~TeV is presented in Fig.~\ref{fig:mbresult}. The distributions are compared to PYTHIA 8 Monash 2013 (PYTHIA version: 8.311) and EPOS4 (EPOS4 version: 4.0.0) for the same event class~\cite{Skands:2014pea, Sjostrand:2014zea, Bierlich:2022pfr, Werner:2023zvo}. 
Different color reconnection modes (CR1, CR2) in PYTHIA 8 Monash 2013 tune are also shown. PYTHIA 8 CR1 and CR2 are alternative color-reconnection variants of the default Monash 2013 tune: CR1 uses a QCD-based reconnection model that can form junctions, while CR2 uses the gluon-move model in which only gluons are reconnected to reduce string length.

PYTHIA 8 Monash 2013, which was developed using LHC minimum-bias and underlying-event data from $\sqrt{s} = 0.9$ to 13 TeV, provides an excellent description of the data across the entire pseudorapidity range. The ratio plot in the lower panel of Fig.~\ref{fig:mbresult} indicates that the model agrees with the measurement within 1\%, confirming that it accurately reproduces both the shape and the normalization of the charged-particle pseudorapidity distribution. 
On the other hand, EPOS4 exhibits deviations from the data, underestimating by $\sim$7\% the measurement at $\eta=0$.
PYTHIA 8 CR2 exhibits a systematically higher prediction, exceeding the data by approximately 4\%, whereas PYTHIA 8 CR1 is lower than the measurement by $\sim$3\%. Despite these differences in normalization, the ratios for both CR1 and CR2 remain relatively flat, indicating that these variations of PYTHIA maintain the overall shape of the pseudorapidity distribution close to the data, but with a different normalization.

\begin{figure}[h!tb]
\centering
\includegraphics[width=0.8\linewidth]{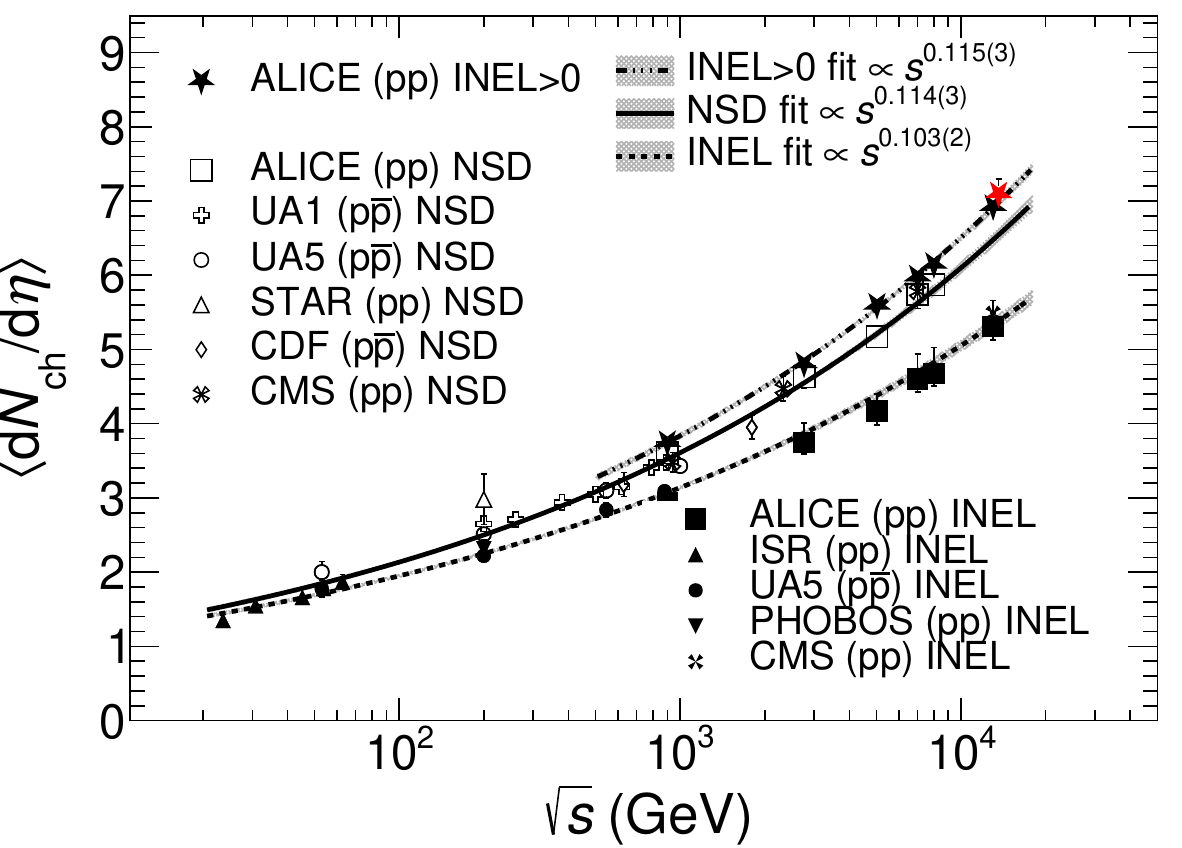}
\caption{The values of $\avdndeta$ in $|\eta|<0.5$ for the INEL, NSD, and \inelg event classes as a function of center-of-mass energy~\cite{ALICE:2009wpl, ALICE:2010cin, ALICE:2010mty, ALICE:2015olq, ALICE:2015qqj, ALICE:2020swj,UA1:1989bou, UA5:1986yef, STAR:2008med, CDF:1989nkn, CMS:2010tjh, CMS:2010wcx, CMS:2015zrm, Ames-Bologna-CERN-Dortmund-Heidelberg-Warsaw:1983cqw, PHOBOS:2004xnp}. The lines depict a power-law fit for each event class. The gray bands show one standard deviation of the fit where the errors are quite small. The $\avdndeta$ value for \inelg\ events in pp collisions at \cmss\ is included (the red star symbol). Note that the newly added data point is not included in the fit. }
\label{fig:energydependence2}
\end{figure}

Figure~\ref{fig:energydependence2} presents the center-of-mass energy dependence of the average charged-particle pseudorapidity density ($\avdndeta$) measured at midrapidity ($|\eta|<0.5$) for different event classes, including inelastic (INEL), non-single diffractive (NSD), and \inelg\ events. 
The measured value of $\avdndeta$ averaged over $|\eta|<0.5$ for \inelg\ events in pp collisions at \cmss\ is  
\begin{equation*}
 \avdndeta = 7.10\pm0.20 .
\end{equation*}
The ALICE measurements, together with previous data from ISR, UA5, UA1, PHOBOS, STAR, CDF, and CMS~\cite{UA1:1989bou, UA5:1986yef, STAR:2008med, CDF:1989nkn, CMS:2010tjh, CMS:2010wcx, CMS:2015zrm, Ames-Bologna-CERN-Dortmund-Heidelberg-Warsaw:1983cqw, PHOBOS:2004xnp} are shown in Fig.~\ref{fig:energydependence2}. 
The measured values of $\avdndeta$ exhibit a clear power-law dependence on the center-of-mass energy and can be parameterized as $\avdndeta \propto s^\delta$. 
The extracted scaling exponent for \inelg\ events is  $\delta = 0.115\pm0.003$ up to $\sqrt{s} = 13$ TeV.
The result presented in this document at \cmss, i.e., the highest energy in pp collisions, remains consistent with the ones for lower center-of-mass energies. If the new data point at $\sqrt{s}=13.6$ TeV is included in the fit, the resulting  $\delta$ value does not change. 
It confirms that charged-particle production at midrapidity follows a power-law scaling behavior across a wide range of collision energies, from low-energy fixed-target experiments to the highest LHC energies. 

\begin{figure}[h!tbp]
\centering
\includegraphics[width=0.78\linewidth]{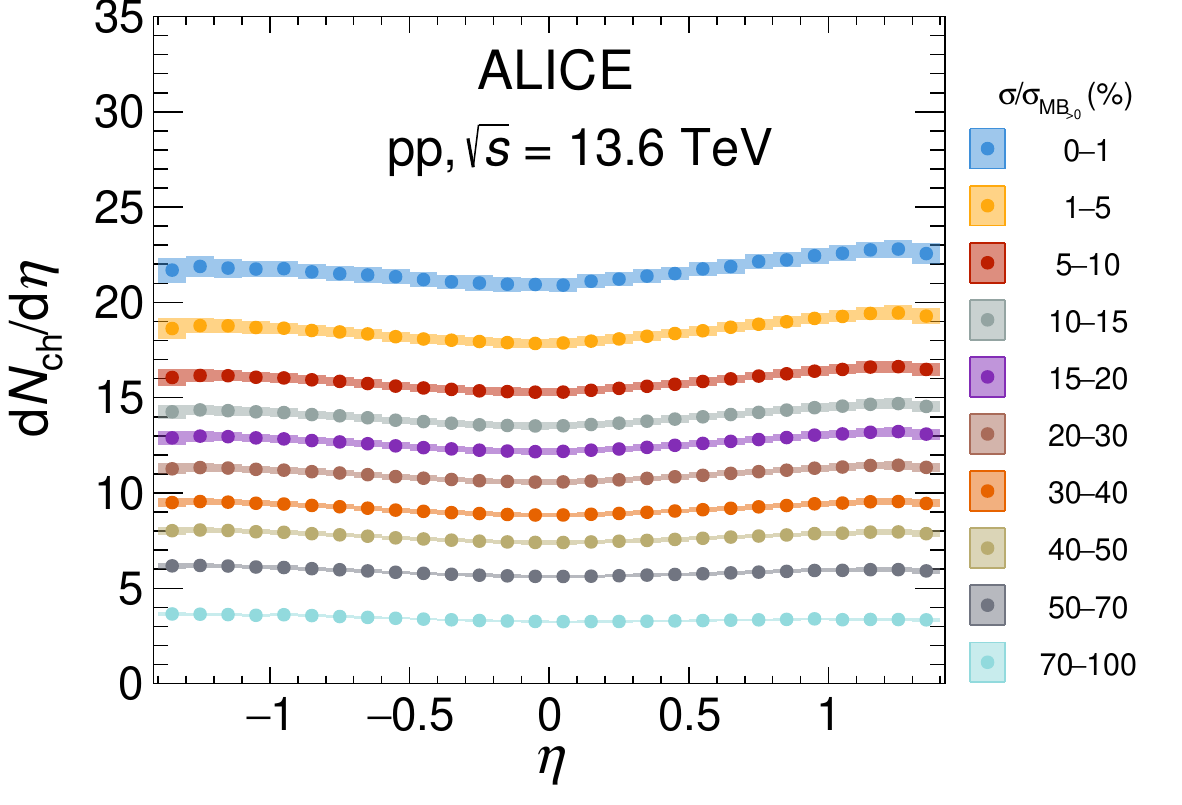}

\caption{Charged-particle pseudorapidity density as a function of $\eta$ for the 0--1 to 70--100\% multiplicity classes measured with the forward multiplicity estimator, illustrating the dependence of particle production on forward event activity. Higher event activity classes exhibit an overall increase in charged-particle multiplicity, while the shape of the distributions remains consistent across different multiplicity classes.}
\label{fig:percentile}
\end{figure}

The distributions of \dndeta\ for different multiplicity classes are shown in Fig.~\ref{fig:percentile}. The multiplicity classification is performed using the forward multiplicity estimator, which is based on signals from the FT0A and FT0C detectors. The results reveal a strong dependence of charged-particle production on event multiplicity.  The highest multiplicity class (0--1\%) shows an increased \dndeta\ value compared to lower-multiplicity events, consistent with the expected correlation between event activity and charged-particle production. The yield in this class is approximately 6.4 times that of the lowest multiplicity class (70--100\%). The asymmetric pseudorapidity acceptance of the FT0 detectors may cause an asymmetry in the distributions of positive and negative pseudorapidities. These results provide valuable input for the development of theoretical models aimed at describing the charged-particle production over a wide pseudorapidity range.

The average $\avdndeta$ values over the pseudorapidity range $|\eta|<0.5$ for different multiplicity classes in pp collisions at \cmss\ are reported in Tab.~\ref{tab:multdependentavdndeta}. 
The table also lists the MB>0 multiplicity classes in data together with the matched INEL>0 percentiles used for the model calculations. 
A decreasing trend is observed in the data with decreasing event multiplicity, with the highest-multiplicity class (0--1\%) exhibiting the largest value of $\avdndeta$, reaching approximately 21, while the lowest-multiplicity class (70--100\%) shows a significantly smaller value of around 3.3.

\begin{table}[hb!]
\centering
\caption{Multiplicity dependent $\avdndeta$ values in $|\eta|<0.5$ in pp collisions at \cmss. Correspondence of the multiplicity classes between MB>0 and INEL>0 is listed together. Model calculations for the corresponding INEL>0 percentiles are also reported. These values represent the estimation for the specific dataset used in this analysis and may vary with detector conditions.}
\label{tab:multdependentavdndeta}
\resizebox{1.0\textwidth}{!}{%
\begin{tabular}{ c | c | c || c | c | c | c }
\hline
\multirow{2}{*}{\shortstack{MB>0\\percentile (\%)}}
& \multirow{2}{*}{\shortstack{INEL>0\\percentile (\%)}}
& \multirow{2}{*}{Data}
& \multicolumn{3}{c|}{PYTHIA 8 Monash}
& \multirow{2}{*}{EPOS4}
\\ \cline{4-6}
& & & Default & CR1 & CR2 &  \\
\hline

0--1    & 0--0.89       & $21.13\pm0.37$ & 26.01 & 23.93 & 31.41 & 27.89 \\
1--5    & 0.89--4.42    & $18.03\pm0.26$ & 21.24 & 19.78 & 25.09 & 21.53 \\
5--10   & 4.42--8.83    & $15.44\pm0.21$ & 17.79 & 16.71 & 20.46 & 17.37 \\
10--15  & 8.83--13.24   & $13.65\pm0.19$ & 15.33 & 14.51 & 17.11 & 14.67 \\
15--20  & 13.24--17.66  & $12.29\pm0.17$ & 13.39 & 12.76 & 14.47 & 12.68 \\
20--30  & 17.66--26.52  & $10.68\pm0.15$ & 10.96 & 10.62 & 11.35 & 10.24 \\
30--40  & 26.52--35.46  & $8.93\pm0.13$  & 8.42  & 8.35  & 8.22  & 7.77  \\
40--50  & 35.46--44.50  & $7.48\pm0.11$  & 6.49  & 6.57  & 6.01  & 5.89  \\
50--70  & 44.50--63.27  & $5.68\pm0.09$  & 4.35  & 4.54  & 3.93  & 3.82  \\
70--100 & 63.27--100    & $3.29\pm0.06$  & 2.58  & 2.62  & 2.44  & 2.20  \\
\hline
0--100  & 0--100        & $7.10\pm0.20$  & 7.11  & 6.94  & 7.37  & 6.64  \\

\hline
\end{tabular}
}
\end{table}

\begin{figure}[h!tbp]
\centering
\includegraphics[width=\linewidth]{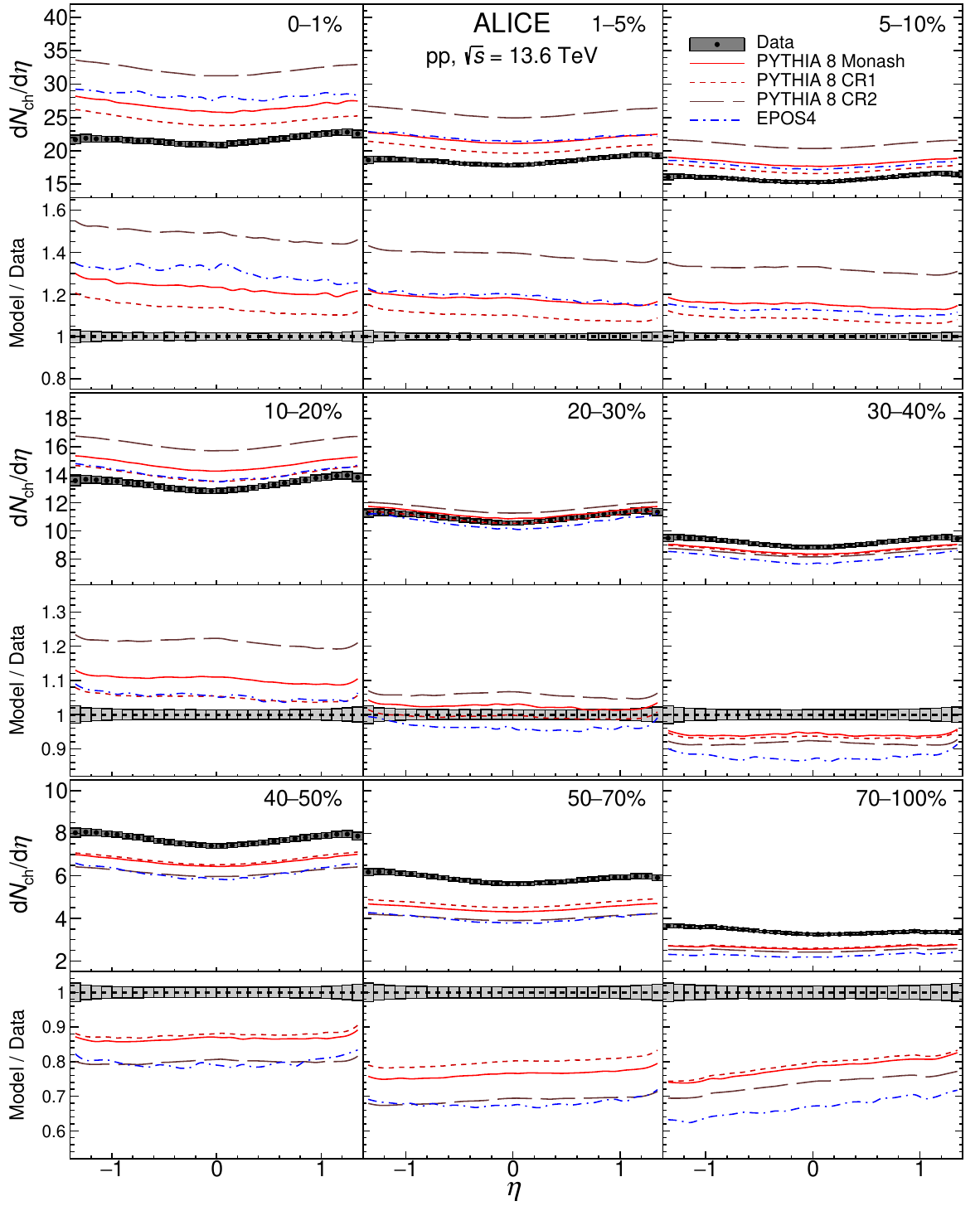}
\caption{\dndeta\ distributions for different multiplicity classes in pp collisions at \cmss. The top, third and fifth rows present \dndeta\ distributions from the 0--1\% to the 70--100\% multiplicity classes, and the other rows display the model-to-data ratios of the rows above. The distributions are compared with theoretical predictions from PYTHIA~8 with their variations and EPOS4. The multiplicity classes in the data are defined using the MB>0 selection, whereas those in the models correspond to the matching \inelg percentiles, as listed in Tab.~\ref{tab:multdependentavdndeta}.}
\label{fig:percentilemodel1}
\end{figure}

The \dndeta\ distributions in pp collisions at \cmss\ for different multiplicity classes are compared with theoretical predictions in Fig.~\ref{fig:percentilemodel1}. 
The top, third and fifth rows present the \dndeta\ distributions of data and the predictions for various multiplicity classes, ranging from the highest (0--1\%) to the lowest (70--100\%). 
The other rows display the model-to-data ratios for their upper panels, quantitatively evaluating the agreement between experimental measurements and theoretical expectations.
In this comparison, the multiplicity classes in the data are defined using the MB>0 selection, while the model results are shown for the matched INEL>0 percentile intervals listed in Tab.~\ref{tab:multdependentavdndeta}.

For events in the highest multiplicity interval (0--1\%), all model predictions overestimate the measured $\dndeta$, with the largest deviation observed for PYTHIA~8 CR2. 
As the multiplicity decreases, the model-to-data ratios gradually approach unity, with the best overall agreement found in the intermediate multiplicity classes, around 20--30\%. 
At lower multiplicities, all models underestimate the data; among the predictions, EPOS4 generally gives lower values than the PYTHIA~8 variations, while PYTHIA~8 CR2 is also on the lower side in several classes. 
At the lowest multiplicities, all models underestimate the data by up to about 40\%.
Overall, none of the models reproduces the full multiplicity dependence quantitatively.

A comparison with previous ALICE measurements performed during Run~2 at
$\sqrt{s}=13$~TeV~\cite{ALICE:2020swj} shows that the multiplicity-dependent $\mathrm{d}N_{\mathrm{ch}}/\mathrm{d}\eta$ values
measured in this analysis are systematically different for the same MB>0 classes. 
To investigate the origin of the differences observed when comparing the present
multiplicity-dependent results to Run~2 measurements, a dedicated model study was
performed using \textsc{PYTHIA}~8. In this study, multiplicity
classes are defined independently using the V0M and FT0M acceptances for the same
INEL$>$0 event sample, allowing for a direct comparison of the resulting event
activity selections. The study shows that the V0M-based classification selects events with higher charged-particle densities at midrapidity than the FT0M-based one by $\sim 15$\% for the 0--1\% of INEL>0 events. In contrast, for low multiplicity classes, the V0M-based classification results in lower midrapidity particle densities compared to the FT0M-based selection by $\sim 15$\%.

\section{Conclusions}

This paper presents a detailed measurement of the pseudorapidity density of primary charged particles (\dndeta) in $|\eta|<1.5$ in proton--proton collisions at a center-of-mass energy of \cmss, based on data from the LHC Run 3. The measurements provide information on charged-particle production at the highest center-of-mass energy reached at the LHC. The results include both inclusive and multiplicity-dependent analyses.
The minimum-bias result is corrected for \inelg\ events, requiring at least one charged particle in the pseudorapidity range $|\eta|<1$. 

The \dndeta\ measurement for \inelg\ events reveals an increase in charged-particle production compared to previous pp collision energies, following the energy scaling trend observed in earlier studies. When combined with data from lower center-of-mass energies, the energy dependence of 
\dndeta\ follows a power-law behavior for INEL>0 events. The inclusive \dndeta\ result is compared with theoretical predictions from PYTHIA 8 and EPOS4. PYTHIA 8 Monash 2013 describes the charged-particle yield well, while EPOS4 shows a discrepancy, particularly in the midrapidity region, suggesting that additional tuning is required. 

The study extends to multiplicity-dependent \dndeta\ analyses, confirming a correlation between event multiplicity and charged-particle production. The models overestimate the \dndeta\ in the high-multiplicity event classes while they underestimate the \dndeta\ in the low-multiplicity event classes. The observed discrepancies between models and data across different multiplicity classes highlight the importance of multiplicity-dependent measurements for improving the accuracy of event generators and their tuning. While both PYTHIA 8 and EPOS4 capture the general trends of \dndeta\ distributions, the observed deviations indicate that further refinements in MPI modeling, color reconnection (CR), and hadronization mechanisms are required. 
These results provide a new reference of \dndeta\ at the LHC, offering essential constraints for theoretical descriptions of charged-particle production, QCD dynamics, and hadronization processes in high-energy hadronic collisions.


\newenvironment{acknowledgement}{\relax}{\relax}
\begin{acknowledgement}
\section*{Acknowledgements}

The ALICE Collaboration would like to thank all its engineers and technicians for their invaluable contributions to the construction of the experiment and the CERN accelerator teams for the outstanding performance of the LHC complex.
The ALICE Collaboration gratefully acknowledges the resources and support provided by all Grid centres and the Worldwide LHC Computing Grid (WLCG) collaboration.
The ALICE Collaboration acknowledges the following funding agencies for their support in building and running the ALICE detector:
A. I. Alikhanyan National Science Laboratory (Yerevan Physics Institute) Foundation (ANSL), State Committee of Science and World Federation of Scientists (WFS), Armenia;
Austrian Academy of Sciences, Austrian Science Fund (FWF): [M 2467-N36] and Nationalstiftung f\"{u}r Forschung, Technologie und Entwicklung, Austria;
Ministry of Communications and High Technologies, National Nuclear Research Center, Azerbaijan;
Rede Nacional de F\'{i}sica de Altas Energias (Renafae), Financiadora de Estudos e Projetos (Finep), Funda\c{c}\~{a}o de Amparo \`{a} Pesquisa do Estado de S\~{a}o Paulo (FAPESP) and The Sao Paulo Research Foundation  (FAPESP), Brazil;
Bulgarian Ministry of Education and Science, within the National Roadmap for Research Infrastructures 2020-2027 (object CERN), Bulgaria;
Ministry of Education of China (MOEC) , Ministry of Science \& Technology of China (MSTC) and National Natural Science Foundation of China (NSFC), China;
Ministry of Science and Education and Croatian Science Foundation, Croatia;
Centro de Aplicaciones Tecnol\'{o}gicas y Desarrollo Nuclear (CEADEN), Cubaenerg\'{\i}a, Cuba;
Ministry of Education, Youth and Sports of the Czech Republic, Czech Republic;
The Danish Council for Independent Research | Natural Sciences, the VILLUM FONDEN and Danish National Research Foundation (DNRF), Denmark;
Helsinki Institute of Physics (HIP), Finland;
Commissariat \`{a} l'Energie Atomique (CEA) and Institut National de Physique Nucl\'{e}aire et de Physique des Particules (IN2P3) and Centre National de la Recherche Scientifique (CNRS), France;
Deutsche Forschungs Gemeinschaft (DFG, German Research Foundation) ``Neutrinos and Dark Matter in Astro- and Particle Physics'' (grant no. SFB 1258) 
 and GSI Helmholtzzentrum f\"{u}r Schwerionenforschung GmbH, Germany;
General Secretariat for Research and Technology, Ministry of Education, Research and Religions, Greece;
National Research, Development and Innovation Office, Hungary;
Department of Atomic Energy Government of India (DAE), Department of Science and Technology, Government of India (DST), University Grants Commission, Government of India (UGC) and Council of Scientific and Industrial Research (CSIR), India;
National Research and Innovation Agency - BRIN, Indonesia;
Istituto Nazionale di Fisica Nucleare (INFN), Italy;
Japanese Ministry of Education, Culture, Sports, Science and Technology (MEXT) and Japan Society for the Promotion of Science (JSPS) KAKENHI, Japan;
Consejo Nacional de Ciencia (CONACYT) y Tecnolog\'{i}a, through Fondo de Cooperaci\'{o}n Internacional en Ciencia y Tecnolog\'{i}a (FONCICYT) and Direcci\'{o}n General de Asuntos del Personal Academico (DGAPA), Mexico;
Nederlandse Organisatie voor Wetenschappelijk Onderzoek (NWO), Netherlands;
The Research Council of Norway, Norway;
Pontificia Universidad Cat\'{o}lica del Per\'{u}, Peru;
Ministry of Science and Higher Education, National Science Centre and WUT ID-UB, Poland;
Korea Institute of Science and Technology Information and National Research Foundation of Korea (NRF), Republic of Korea;
Ministry of Education and Scientific Research, Institute of Atomic Physics, Ministry of Research and Innovation and Institute of Atomic Physics and Universitatea Nationala de Stiinta si Tehnologie Politehnica Bucuresti, Romania;
Ministerstvo skolstva, vyskumu, vyvoja a mladeze SR, Slovakia;
National Research Foundation of South Africa, South Africa;
Swedish Research Council (VR) and Knut \& Alice Wallenberg Foundation (KAW), Sweden;
European Organization for Nuclear Research, Switzerland;
Suranaree University of Technology (SUT), National Science and Technology Development Agency (NSTDA) and National Science, Research and Innovation Fund (NSRF via PMU-B B05F650021), Thailand;
Turkish Energy, Nuclear and Mineral Research Agency (TENMAK), Turkey;
National Academy of  Sciences of Ukraine, Ukraine;
Science and Technology Facilities Council (STFC), United Kingdom;
National Science Foundation of the United States of America (NSF) and United States Department of Energy, Office of Nuclear Physics (DOE NP), United States of America.
In addition, individual groups or members have received support from:
Czech Science Foundation (grant no. 23-07499S), Czech Republic;
FORTE project, reg.\ no.\ CZ.02.01.01/00/22\_008/0004632, Czech Republic, co-funded by the European Union, Czech Republic;
European Research Council (grant no. 950692), European Union;
FAIR - Future Artificial Intelligence Research, funded by the NextGenerationEU program (Italy).

\end{acknowledgement}

\bibliographystyle{utphys}   
\bibliography{bibliography}

\newpage
\appendix

%
%

\section{The ALICE Collaboration}
\label{app:collab}
\begin{flushleft} 
\small

I.J.~Abualrob\,\orcidlink{0009-0005-3519-5631}\,$^{\rm 113}$, 
S.~Acharya\,\orcidlink{0000-0002-9213-5329}\,$^{\rm 49}$, 
G.~Aglieri Rinella\,\orcidlink{0000-0002-9611-3696}\,$^{\rm 32}$, 
L.~Aglietta\,\orcidlink{0009-0003-0763-6802}\,$^{\rm 24}$, 
N.~Agrawal\,\orcidlink{0000-0003-0348-9836}\,$^{\rm 25}$, 
Z.~Ahammed\,\orcidlink{0000-0001-5241-7412}\,$^{\rm 133}$, 
S.~Ahmad\,\orcidlink{0000-0003-0497-5705}\,$^{\rm 15}$, 
I.~Ahuja\,\orcidlink{0000-0002-4417-1392}\,$^{\rm 36}$, 
Z.~Akbar$^{\rm 80}$, 
V.~Akishina\,\orcidlink{0009-0004-4802-2089}\,$^{\rm 38}$
% ...

\section*{Affiliation Notes}

$^{\rm I}$ Deceased\\
$^{\rm II}$ Also at: Fondazione Bruno Kessler (FBK), Trento, Italy\\
$^{\rm III}$ Also at: Max-Planck-Institut fur Physik, Munich, Germany\\
$^{\rm IV}$ Also at: Czech Technical University in Prague (CZ)\\
$^{\rm V}$ Also at: Instituto de Fisica da Universidade de Sao Paulo\\
$^{\rm VI}$ Also at: Dipartimento DET del Politecnico di Torino, Turin, Italy\\
$^{\rm VII}$ Also at: Department of Applied Physics, Aligarh Muslim University, Aligarh, India\\
$^{\rm VIII}$ Also at: Institute of Theoretical Physics, University of Wroclaw, Poland\\
$^{\rm IX}$ Also at: Facultad de Ciencias, Universidad Nacional Aut\'{o}noma de M\'{e}xico, Mexico City, Mexico\\

\section*{Collaboration Institutes}

$^{1}$ A.I. Alikhanyan National Science Laboratory (Yerevan Physics Institute) Foundation, Yerevan, Armenia\\
$^{2}$ AGH University of Krakow, Cracow, Poland\\
$^{3}$ Bogolyubov Institute for Theoretical Physics, National Academy of Sciences of Ukraine, Kyiv, Ukraine\\
$^{4}$ Bose Institute, Department of Physics  and Centre for Astroparticle Physics and Space Science (CAPSS), Kolkata, India\\
$^{5}$ California Polytechnic State University, San Luis Obispo, California, United States\\
$^{6}$ Central China Normal University, Wuhan, China\\
$^{7}$ Centro de Aplicaciones Tecnol\'{o}gicas y Desarrollo Nuclear (CEADEN), Havana, Cuba\\
$^{8}$ Centro de Investigaci\'{o}n y de Estudios Avanzados (CINVESTAV), Mexico City and M\'{e}rida, Mexico\\
$^{9}$ Chicago State University, Chicago, Illinois, United States\\
$^{10}$ China Nuclear Data Center, China Institute of Atomic Energy, Beijing, China\\
$^{11}$ China University of Geosciences, Wuhan, China\\
$^{12}$ Chungbuk National University, Cheongju, Republic of Korea\\
$^{13}$ Comenius University Bratislava, Faculty of Mathematics, Physics and Informatics, Bratislava, Slovak Republic\\
$^{14}$ Creighton University, Omaha, Nebraska, United States\\
$^{15}$ Department of Physics, Aligarh Muslim University, Aligarh, India\\
$^{16}$ Department of Physics, Pusan National University, Pusan, Republic of Korea\\
$^{17}$ Department of Physics, Sejong University, Seoul, Republic of Korea\\
$^{18}$ Department of Physics, University of California, Berkeley, California, United States\\
$^{19}$ Department of Physics, University of Oslo, Oslo, Norway\\
$^{20}$ Department of Physics and Technology, University of Bergen, Bergen, Norway\\
$^{21}$ Dipartimento di Fisica, Universit\`{a} di Pavia, Pavia, Italy\\
$^{22}$ Dipartimento di Fisica dell'Universit\`{a} and Sezione INFN, Cagliari, Italy\\
$^{23}$ Dipartimento di Fisica dell'Universit\`{a} and Sezione INFN, Trieste, Italy\\
$^{24}$ Dipartimento di Fisica dell'Universit\`{a} and Sezione INFN, Turin, Italy\\
$^{25}$ Dipartimento di Fisica e Astronomia dell'Universit\`{a} and Sezione INFN, Bologna, Italy\\
$^{26}$ Dipartimento di Fisica e Astronomia dell'Universit\`{a} and Sezione INFN, Catania, Italy\\
$^{27}$ Dipartimento di Fisica e Astronomia dell'Universit\`{a} and Sezione INFN, Padova, Italy\\
$^{28}$ Dipartimento di Fisica `E.R.~Caianiello' dell'Universit\`{a} and Gruppo Collegato INFN, Salerno, Italy\\
$^{29}$ Dipartimento DISAT del Politecnico and Sezione INFN, Turin, Italy\\
$^{30}$ Dipartimento di Scienze MIFT, Universit\`{a} di Messina, Messina, Italy\\
$^{31}$ Dipartimento Interateneo di Fisica `M.~Merlin' and Sezione INFN, Bari, Italy\\
$^{32}$ European Organization for Nuclear Research (CERN), Geneva, Switzerland\\
$^{33}$ Faculty of Electrical Engineering, Mechanical Engineering and Naval Architecture, University of Split, Split, Croatia\\
$^{34}$ Faculty of Nuclear Sciences and Physical Engineering, Czech Technical University in Prague, Prague, Czech Republic\\
$^{35}$ Faculty of Physics, Sofia University, Sofia, Bulgaria\\
$^{36}$ Faculty of Science, P.J.~\v{S}af\'{a}rik University, Ko\v{s}ice, Slovak Republic\\
$^{37}$ Faculty of Technology, Environmental and Social Sciences, Bergen, Norway\\
$^{38}$ Frankfurt Institute for Advanced Studies, Johann Wolfgang Goethe-Universit\"{a}t Frankfurt, Frankfurt, Germany\\
$^{39}$ Fudan University, Shanghai, China\\
$^{40}$ Gauhati University, Department of Physics, Guwahati, India\\
$^{41}$ Helmholtz-Institut f\"{u}r Strahlen- und Kernphysik, Rheinische Friedrich-Wilhelms-Universit\"{a}t Bonn, Bonn, Germany\\
$^{42}$ Helsinki Institute of Physics (HIP), Helsinki, Finland\\
$^{43}$ High Energy Physics Group,  Universidad Aut\'{o}noma de Puebla, Puebla, Mexico\\
$^{44}$ Horia Hulubei National Institute of Physics and Nuclear Engineering, Bucharest, Romania\\
$^{45}$ HUN-REN Wigner Research Centre for Physics, Budapest, Hungary\\
$^{46}$ Indian Institute of Technology Bombay (IIT), Mumbai, India\\
$^{47}$ Indian Institute of Technology Indore, Indore, India\\
$^{48}$ INFN, Laboratori Nazionali di Frascati, Frascati, Italy\\
$^{49}$ INFN, Sezione di Bari, Bari, Italy\\
$^{50}$ INFN, Sezione di Bologna, Bologna, Italy\\
$^{51}$ INFN, Sezione di Cagliari, Cagliari, Italy\\
$^{52}$ INFN, Sezione di Catania, Catania, Italy\\
$^{53}$ INFN, Sezione di Padova, Padova, Italy\\
$^{54}$ INFN, Sezione di Pavia, Pavia, Italy\\
$^{55}$ INFN, Sezione di Torino, Turin, Italy\\
$^{56}$ INFN, Sezione di Trieste, Trieste, Italy\\
$^{57}$ Inha University, Incheon, Republic of Korea\\
$^{58}$ Institute for Gravitational and Subatomic Physics (GRASP), Utrecht University/Nikhef, Utrecht, Netherlands\\
$^{59}$ Institute of Experimental Physics, Slovak Academy of Sciences, Ko\v{s}ice, Slovak Republic\\
$^{60}$ Institute of Physics, Homi Bhabha National Institute, Bhubaneswar, India\\
$^{61}$ Institute of Physics of the Czech Academy of Sciences, Prague, Czech Republic\\
$^{62}$ Institute of Space Science (ISS), Bucharest, Romania\\
$^{63}$ Institut f\"{u}r Kernphysik, Johann Wolfgang Goethe-Universit\"{a}t Frankfurt, Frankfurt, Germany\\
$^{64}$ Instituto de Ciencias Nucleares, Universidad Nacional Aut\'{o}noma de M\'{e}xico, Mexico City, Mexico\\
$^{65}$ Instituto de F\'{i}sica, Universidade Federal do Rio Grande do Sul (UFRGS), Porto Alegre, Brazil\\
$^{66}$ Instituto de F\'{\i}sica, Universidad Nacional Aut\'{o}noma de M\'{e}xico, Mexico City, Mexico\\
$^{67}$ iThemba LABS, National Research Foundation, Somerset West, South Africa\\
$^{68}$ Jeonbuk National University, Jeonju, Republic of Korea\\
$^{69}$ Korea Institute of Science and Technology Information, Daejeon, Republic of Korea\\
$^{70}$ Laboratoire de Physique Subatomique et de Cosmologie, Universit\'{e} Grenoble-Alpes, CNRS-IN2P3, Grenoble, France\\
$^{71}$ Lawrence Berkeley National Laboratory, Berkeley, California, United States\\
$^{72}$ Lund University Department of Physics, Division of Particle Physics, Lund, Sweden\\
$^{73}$ Marietta Blau Institute, Vienna, Austria\\
$^{74}$ Nagasaki Institute of Applied Science, Nagasaki, Japan\\
$^{75}$ Nara Women{'}s University (NWU), Nara, Japan\\
$^{76}$ National and Kapodistrian University of Athens, School of Science, Department of Physics , Athens, Greece\\
$^{77}$ National Centre for Nuclear Research, Warsaw, Poland\\
$^{78}$ National Institute of Science Education and Research, Homi Bhabha National Institute, Jatni, India\\
$^{79}$ National Nuclear Research Center, Baku, Azerbaijan\\
$^{80}$ National Research and Innovation Agency - BRIN, Jakarta, Indonesia\\
$^{81}$ Niels Bohr Institute, University of Copenhagen, Copenhagen, Denmark\\
$^{82}$ Nikhef, National institute for subatomic physics, Amsterdam, Netherlands\\
$^{83}$ Nuclear Physics Group, STFC Daresbury Laboratory, Daresbury, United Kingdom\\
$^{84}$ Nuclear Physics Institute of the Czech Academy of Sciences, Husinec-\v{R}e\v{z}, Czech Republic\\
$^{85}$ Oak Ridge National Laboratory, Oak Ridge, Tennessee, United States\\
$^{86}$ Ohio State University, Columbus, Ohio, United States\\
$^{87}$ Physics department, Faculty of science, University of Zagreb, Zagreb, Croatia\\
$^{88}$ Physics Department, Panjab University, Chandigarh, India\\
$^{89}$ Physics Department, University of Jammu, Jammu, India\\
$^{90}$ Physics Program and International Institute for Sustainability with Knotted Chiral Meta Matter (WPI-SKCM$^{2}$), Hiroshima University, Hiroshima, Japan\\
$^{91}$ Physikalisches Institut, Eberhard-Karls-Universit\"{a}t T\"{u}bingen, T\"{u}bingen, Germany\\
$^{92}$ Physikalisches Institut, Ruprecht-Karls-Universit\"{a}t Heidelberg, Heidelberg, Germany\\
$^{93}$ Physik Department, Technische Universit\"{a}t M\"{u}nchen, Munich, Germany\\
$^{94}$ Politecnico di Bari and Sezione INFN, Bari, Italy\\
$^{95}$ Research Division and ExtreMe Matter Institute EMMI, GSI Helmholtzzentrum f\"ur Schwerionenforschung GmbH, Darmstadt, Germany\\
$^{96}$ Saga University, Saga, Japan\\
$^{97}$ Saha Institute of Nuclear Physics, Homi Bhabha National Institute, Kolkata, India\\
$^{98}$ School of Physics and Astronomy, University of Birmingham, Birmingham, United Kingdom\\
$^{99}$ Secci\'{o}n F\'{\i}sica, Departamento de Ciencias, Pontificia Universidad Cat\'{o}lica del Per\'{u}, Lima, Peru\\
$^{100}$ SUBATECH, IMT Atlantique, Nantes Universit\'{e}, CNRS-IN2P3, Nantes, France\\
$^{101}$ Sungkyunkwan University, Suwon City, Republic of Korea\\
$^{102}$ Suranaree University of Technology, Nakhon Ratchasima, Thailand\\
$^{103}$ Technical University of Ko\v{s}ice, Ko\v{s}ice, Slovak Republic\\
$^{104}$ The Henryk Niewodniczanski Institute of Nuclear Physics, Polish Academy of Sciences, Cracow, Poland\\
$^{105}$ The University of Texas at Austin, Austin, Texas, United States\\
$^{106}$ Universidad Aut\'{o}noma de Sinaloa, Culiac\'{a}n, Mexico\\
$^{107}$ Universidade de S\~{a}o Paulo (USP), S\~{a}o Paulo, Brazil\\
$^{108}$ Universidade Estadual de Campinas (UNICAMP), Campinas, Brazil\\
$^{109}$ Universidade Federal do ABC, Santo Andre, Brazil\\
$^{110}$ Universitatea Nationala de Stiinta si Tehnologie Politehnica Bucuresti, Bucharest, Romania\\
$^{111}$ University of Cape Town, Cape Town, South Africa\\
$^{112}$ University of Derby, Derby, United Kingdom\\
$^{113}$ University of Houston, Houston, Texas, United States\\
$^{114}$ University of Jyv\"{a}skyl\"{a}, Jyv\"{a}skyl\"{a}, Finland\\
$^{115}$ University of Kansas, Lawrence, Kansas, United States\\
$^{116}$ University of Liverpool, Liverpool, United Kingdom\\
$^{117}$ University of Science and Technology of China, Hefei, China\\
$^{118}$ University of Silesia in Katowice, Katowice, Poland\\
$^{119}$ University of South-Eastern Norway, Kongsberg, Norway\\
$^{120}$ University of Tennessee, Knoxville, Tennessee, United States\\
$^{121}$ University of the Witwatersrand, Johannesburg, South Africa\\
$^{122}$ University of Tokyo, Tokyo, Japan\\
$^{123}$ University of Tsukuba, Tsukuba, Japan\\
$^{124}$ Universit\"{a}t M\"{u}nster, Institut f\"{u}r Kernphysik, M\"{u}nster, Germany\\
$^{125}$ Universit\'{e} Clermont Auvergne, CNRS/IN2P3, LPC, Clermont-Ferrand, France\\
$^{126}$ Universit\'{e} de Lyon, CNRS/IN2P3, Institut de Physique des 2 Infinis de Lyon, Lyon, France\\
$^{127}$ Universit\'{e} de Strasbourg, CNRS, IPHC UMR 7178, F-67000 Strasbourg, France, Strasbourg, France\\
$^{128}$ Universit\'{e} Paris-Saclay, Centre d'Etudes de Saclay (CEA), IRFU, D\'{e}partment de Physique Nucl\'{e}aire (DPhN), Saclay, France\\
$^{129}$ Universit\'{e}  Paris-Saclay, CNRS/IN2P3, IJCLab, Orsay, France\\
$^{130}$ Universit\`{a} degli Studi di Foggia, Foggia, Italy\\
$^{131}$ Universit\`{a} del Piemonte Orientale, Vercelli, Italy\\
$^{132}$ Universit\`{a} di Brescia, Brescia, Italy\\
$^{133}$ Variable Energy Cyclotron Centre, Homi Bhabha National Institute, Kolkata, India\\
$^{134}$ Warsaw University of Technology, Warsaw, Poland\\
$^{135}$ Wayne State University, Detroit, Michigan, United States\\
$^{136}$ Yale University, New Haven, Connecticut, United States\\
$^{137}$ Yildiz Technical University, Istanbul, Turkey\\
$^{138}$ Yonsei University, Seoul, Republic of Korea\\
$^{139}$ Affiliated with an institute formerly covered by a cooperation agreement with CERN\\
$^{140}$ Affiliated with an international laboratory covered by a cooperation agreement with CERN.\\

\end{flushleft} 
  
\end{document}